# Controllable ultra-broadband supercontinuum during two-color femtosecond laser filamentation in high-pressure gases

**Yury E. Geints[1,*], Victor O. Kompanets[2,†], Sergey V. Chekalin[2,†]**

[1]*V. E. Zuev Institute of Atmospheric Optics SB RAS, Acad. Zuev Square 1, Tomsk 634055, Russia*
[2]*Institute of spectroscopy RAS, Fizicheskaya Str. 5, Troitsk, Moscow 108840, Russia*
[†]*These authors contributed equally.*
[*]*Corresponding author e-mail: ygeints@iao.ru*

**Abstract**. Two-color laser filamentation is considered a promising strategy for deep transformations of the spectrum and generating ultrashort light pulses, which is critically important for developing effective methods to control broadband coherent radiation in attosecond physics and ultrafast spectroscopy. We present the results of our experiments on the study of the spectral dynamics of supercontinuum generated during collinear two-color filamentation of femtosecond pulses at the fundamental (800 nm) and second harmonic (400 nm) frequencies in Ar, $N_2$, and $CO_2$ gases at pressures up to 11 atm. The practical importance of the work lies in the systematic analysis of the regime of significant exceeding of the filamentation threshold, achieved not by increasing the energy of the pulses, but by increasing the gas pressure, which made it possible to avoid the limitations associated with the radiation resistance of nonlinear second-harmonic crystals. We demonstrate that the transition to multiple filamentation in high-pressure gases qualitatively changes the interaction dynamics between the color components. We demonstrate that instead of a simple merging of spectra broadened during filamentation, a complex, non-monotonic dependence of the shape of the two-color supercontinuum on the time delay between pulses is observed, including selective suppression of radiation in the 400 nm region. For the first time, significant differences are identified between the dynamics of two-color filamentation in atomic and molecular gases. Particularly, in $CO_2$, spectral transformation is preserved at picosecond interpulse delays, which is explained by the strong contribution of molecular rotational wave packets to the effective nonlinear polarizability. Our results demonstrate the possibility of controlled redistribution of spectral energy and dynamic "switching off" of individual regions of supercontinuum in atomic and molecular gases as their pressure changes, which fills the gap in understanding the spatio-temporal evolution of multiple two-color filaments and paves the way for the controlled formation of ultrashort laser pulses.

## 1. Introduction

Two-color laser filamentation (also referred to as bi-color filamentation) is a nonlinear propagation regime in which two high-power ultrashort laser pulses with different central wavelengths are launched collinearly (or non-collinearly) into a nonlinear medium, either simultaneously or with a certain time delay [1]. The interaction between the pulses occurs through a variety of nonlinear effects. Primarily, when the pulses overlap spatially, four-wave mixing (FWM) [2] and cross-phase modulation (XPM) of the two-color optical fields [3] generate new spectral components and create a cross-Kerr lens. In this process, the field of one pulse imparts a nonlinear phase onto the other, and vice versa, and a strong dependence on the polarization state of each overlapping wave is observed, leading to either suppression or enhancement of the cross-components. Other important physical mechanisms of two-color filamentation involve temporally nonlocal interactions between the components through distinctive wakes left by the pulses in the propagation medium, caused by multiphoton ionization of the medium (plasma wake [1]) and orientationally induced changes in molecular polarizability (quantum wake [4, 5]), since the refractive-index modification differs for molecules aligned parallel or perpendicular to the laser polarization vector. This complex dynamic interaction gives rise to effects that are inaccessible in single-color filamentation, thus enabling control over the supercontinuum (SC) generated during the filamentation process.

An analysis of the most relevant publications on this topic [1-8] shows that the primary outcome of this spectral management is the combination of spectra from separate pulses. Instead of two separate spectra from the "red" (1ω-pulse, e.g., at 800 nm) and the "blue" (2ω-pulse, e.g., at 400 nm) pulses, their co-propagation and interaction in the medium with a certain time delay $\Delta t$ results in a single continuous spectrum spanning two or more octaves (from ~200 nm to ~900 nm). The optimal time delay $\Delta t^*$ can vary depending on the medium and the radiation parameters ($\Delta t^* = 0$ in argon [6], and $\Delta t^* = -300\ldots400$ fs in sapphire [8]) due to the large difference in group-velocity dispersion (GVD) between the harmonics.

Importantly, by varying the delay between the pulses one can manipulate the resulting supercontinuum spectrum. For example, one can selectively enhance the red or blue part of the blue pulse spectrum because cross-phase modulation acts on the leading or trailing edge of the pulse, shifting the frequency toward the red or blue region, respectively. Moreover, even relatively small changes in the delay, on the order of tenths of the pulse duration [8], can result in either complete suppression or restoration of entire portions of the supercontinuum. Clearly, the dynamic spectral control depends on which pulse arrives first at the filamentation region. It has been established that at negative delays, when the IR pulse (1ω-pulse) arrives first, the strongest interaction between the two-color filaments occurs, which can lead to complete quenching of the supercontinuum in the UV range and modulation in the IR region. This is associated, among other factors, with the larger diameter of the ionized channels formed in the plasma wake of the 1ω-pulse.

Furthermore, several studies [1, 8] report the observation of a peculiar synergistic effect from the interaction of two filaments. When the power of each individual pulse is below the threshold for single-filament formation, their joint propagation at certain delays leads to the generation of supercontinuum characteristic of filamentation. In other words, the individual pulses help each other to focus and ionize the medium, which is experimentally confirmed by changes in the nonlinear focus position and the intensity of plasma luminescence of the filament. Therefore, two-color filamentation serves as a potent instrument for the active manipulation of supercontinuum properties. This method demonstrates the ability for extremely fast (on the scale of fractions of the pulse duration) spectral control, up to complete "switching on" and "switching off" of individual spectral regions, and also allows significant broadening of the emission spectrum [2] and the generation of ultrashort pulses without additional compression, which is critically important in attosecond physics [6].

At the same time, all studies known to date have dealt with pulses of relatively low power, close to the single-filamentation threshold in the medium, which is typically limited by the low optical damage threshold of the nonlinear second-harmonic crystals (usually a BBO crystal) used for frequency conversion. Therefore, it is crucial to investigate the effect of collinear two-color filamentation at powers that substantially exceed this threshold (non-collinear interaction of two-color pulses was examined, for instance, in [9]). This transition shifts the process from single-filament to multi-filament mode, complicating the filamentation dynamics and potentially introducing new characteristics into the supercontinuum generation process. To this end, given the power limitations imposed by the second-harmonic crystal, it is advisable not to increase the pulse energy but rather to change the propagation medium itself, making it denser by using optical cells with compressed gases. According to the pressure scaling laws of laser filamentation [10, 11], increasing the gas pressure is equivalent to a proportional increase in the pulse power that would propagate at normal pressure. Therefore, this is the strategy we follow in this study and present the results of our experiments on manipulating the supercontinuum spectrum of two-color filamentation in three gases as their pressure changes.

## 2. Experimental methodic and results

For the experimental investigation of two-color filamentation, we employed collinear interaction of pulses at the fundamental frequency 1ω (central wavelength 800 nm) and the second harmonic 2ω (wavelength 400 nm). The experimental setup is shown in Fig. 1(a). Optical radiation from a regenerative Ti:sapphire amplifier (2.5 mJ pulse energy, 20 Hz repetition rate, 45 fs duration, 8 mm $1/e^2$ beam diameter) was directed onto a 0.25-mm-thick BBO nonlinear crystal to generate the second harmonic. A dichroic beam splitter reflected the 2ω radiation to the side and after polarization rotation by a half-wave plate it was recombined into a collinear geometry. The fundamental beam transmitted through the splitter was sent to a delay line to equalize the optical paths. The energy of the resulting 2ω pulse at the gas-cell entrance was about 500 μJ.

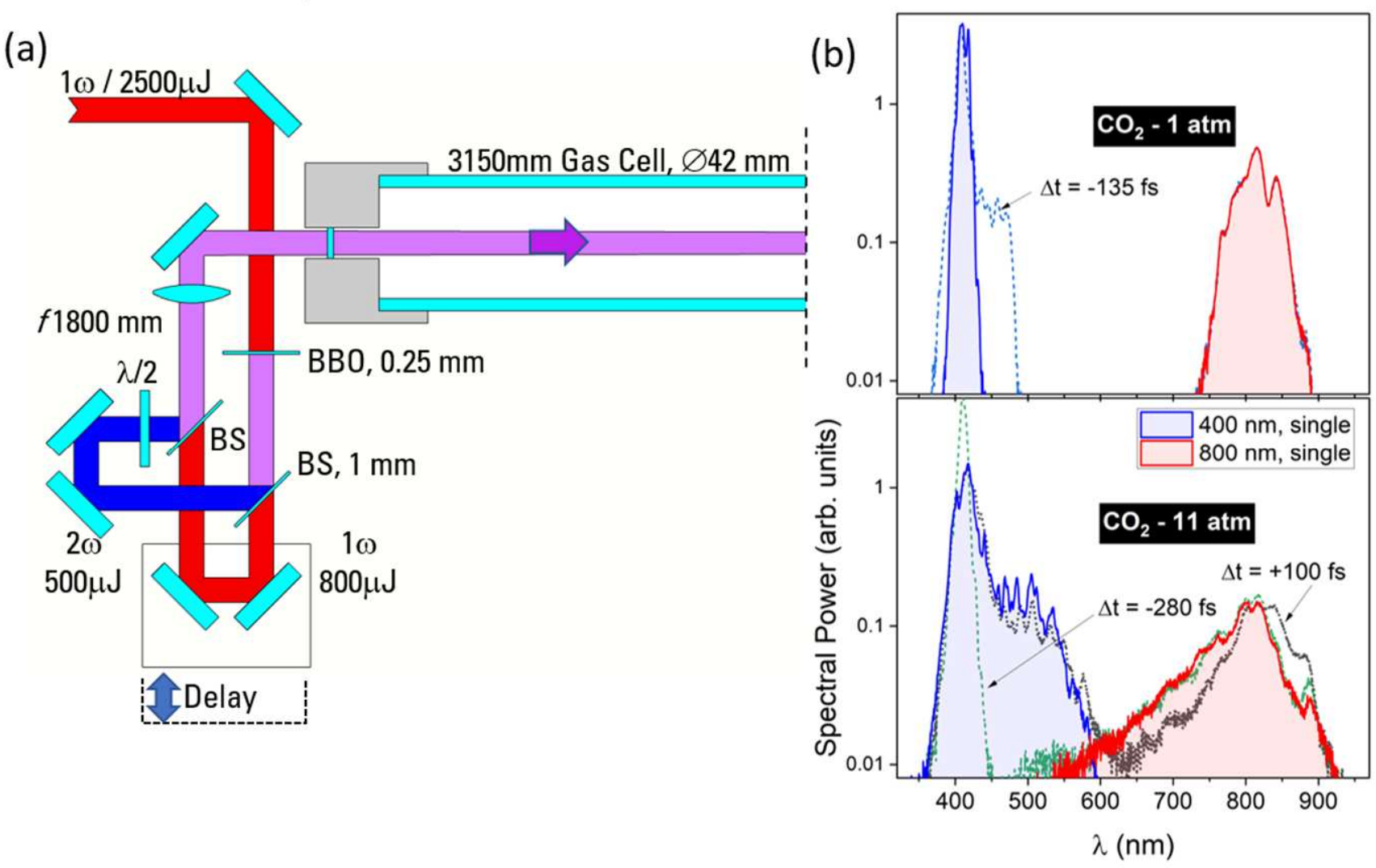

Fig. 1. Experimental setup and representative spectra of two-color filamentation in compressed gas. (a) Schematic of the generation and collinear launching of 1ω (800 nm, 800 μJ) and 2ω (400 nm, 500 μJ) pulses into a 3150-mm-long gas cell; BS – beam splitters, λ/2 – half-wave plate, BBO –nonlinear crystal, *f* – focusing lens. (b) Supercontinuum spectra after filamentation in $CO_2$ at atmospheric (1 atm) and high (11 atm) pressures for different time delays $\Delta t$ of the 2ω pulse, where $\Delta t = 0$ corresponds to spatiotemporal overlap of the pulses. For comparison, spectra obtained from single-pulse propagation at 400 and 800 nm are also shown (shaded regions).

The relative time delay between the two pulses, $\Delta t$, was adjusted by an optical delay line. Thus, together with the gas pressure, $\Delta t$ served as one of the main experimental parameters. The combined two-color beam was focused by a lens of focal length $f$ = 1800 mm into an optical cell filled with various gases. The condition $\Delta t = 0$ was defined as nominal temporal overlap, taking into account the chromatic dispersion of the lens and the difference in pulse energies at the two frequencies. For each gas, we ensured spatial overlap of the plasma filaments at both frequencies in the focal region. The cell was a thick-walled polycarbonate tube with 36-mm inner diameter, 3150-mm length, and 3-mm wall thickness, rated for working pressures up to 15 atm. The entrance and exit windows were made of fused silica with thicknesses of 2 mm and 6 mm, respectively. We investigated the spectral transformations of the femtosecond laser pulses in three gases – atomic argon (Ar), as well as molecular nitrogen ($N_2$) and carbon dioxide ($CO_2$). This choice was motivated by the substantial differences in their internal structure, which lead to distinct nonlinear optical responses under intense laser radiation. All three gases have close values of the electronic Kerr nonlinearity in the near-IR range [12], but they exhibit drastically different Raman-induced spectral broadening during optical-pulse filamentation, being maximal in $CO_2$ and completely absent in atomic argon [13]. The use of compressed gases allowed us to increase the medium nonlinearity without increasing the laser pulse energy. According to the density scaling of the nonlinear response, an increase in pressure is, to first order, equivalent to an increase in the effective optical power for propagation in a gas at normal pressure.

As an example, Fig. 1(b) shows the pulse spectra after propagation through the filamentation region in $CO_2$ at pressures of 1 and 11 atm. For comparison, the spectra obtained from single pulses at 400 and 800 nm are also shown. At atmospheric pressure, the spectra of the individual pulses remain largely isolated, whereas their co-propagation leads to a substantial modification of the spectral distribution. In particular, at a delay $\Delta t = -135$ fs, when the 2ω harmonic lags behind by almost the entire pulse duration, new spectral components appear in the intermediate region between the pulses. This indicates a strong nonlinear action of the IR pulse on the wakeing 2ω pulse through the ionized regions in the filamentation zone.

The character of the spectral modification changes significantly when $CO_2$ pressure is increased to 11 atm. For delays $\Delta t = -280$ and +100 fs, a broad joint spectrum is formed, covering both the fundamental and second-harmonic regions. The spectral structure is markedly different from that observed at atmospheric pressure, as will be discussed in detail below. Thus, increasing the density of the nonlinear medium not only enhances the nonlinear self-action of each pulse [14] but also qualitatively changes the supercontinuum generation dynamics.

The obtained results should be considered in the context of the dynamic interaction of two filaments. When two pulses propagate, their fields interact via cross-phase modulation if the delay does not exceed the pulse duration, as well as through the quantum and plasma wakes generated by rotational rearrangement and ionization of the medium. Therefore, varying the time delay determines which pulse enters the nonlinear interaction region first and how the frequency content of the second pulse is modified. In particular, for negative delays, when the 1ω pulse arrives before the 2ω pulse, the strongest interaction between the two-color filaments is expected. Previous publications have shown that such a configuration can lead to pronounced modulation of the supercontinuum, including suppression of certain spectral regions.

## 3. Discussion

Of particular interest is the pressure dependence of the two-color filamentation. Figure 2 presents supercontinuum spectrograms as functions of gas pressure and time delay between the pulses. Such a dataset allows one to disentangle several factors governing the two-pulse interaction dynamics: the time delay, the density of the nonlinear medium, and its atomic/molecular properties. Each spectrogram was obtained by combining individual pulse spectra normalized to the maximum near the 1ω and 2ω spectral lines. In other words, the continuous supercontinuum spectrum was split into two separate spectra at an artificial boundary of 600 nm, and each part was normalized to its own maximum. These normalized spectral parts were then recombined into a single spectrum, which enabled better visualization of the dynamics of low-intensity spectral components.

It is evident that the presented spectra reveal a substantial transformation of the two-color filamentation dynamics with increasing gas pressure. At atmospheric pressure, in all investigated media, the spectrum retains a predominantly two-component structure corresponding to the original pulses at 400 and 800 nm, and its dependence on the time delay is relatively weak. Increasing the pressure to 6 atm leads to a marked enhancement of nonlinear pulse interaction and pronounced spectral broadening, most noticeable for the 2ω component. When the pressure is raised to 11 atm, the spectral dynamics become much more complicated. In addition to broadening, areas of selective spectral suppression emerge, and their position is determined by the time delay between the pulses.

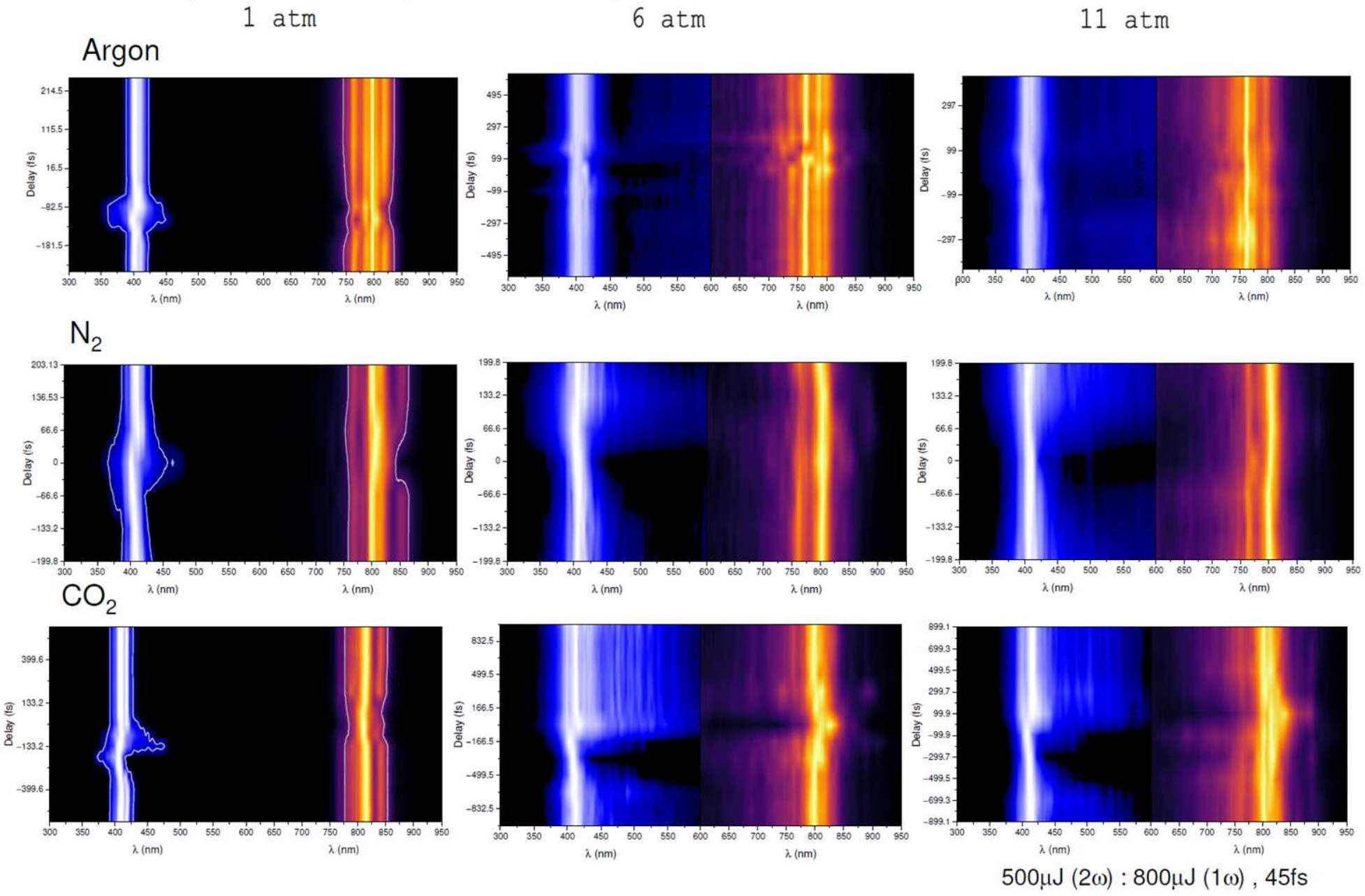

Fig. 2. Supercontinuum spectrograms (normalized to the maximum) for two-color filamentation of 2ω (400 nm, “blue”) and 1ω (800 nm, “red”) pulses in Ar, $N_2$, and $CO_2$ (rows) at pressures of 1, 6, and 11 atm (columns). Negative and positive time delays correspond to the 1ω pulse arriving before and after the 2ω pulse, respectively. Pulse energies: $E_{2\omega} = 500$ μJ and $E_{1\omega} = 800$ μJ, pulse duration is 45 fs.

The observed behavior arises from the combined action of cross-phase modulation, self-phase modulation, and the plasma contribution to the nonlinear refraction. The sign and magnitude of the delay determine the sequence in which the pulses traverse the nonlinear medium and, consequently, the nature of the influence of the plasma wake and the nonlinear phase generated by the first pulse in the interaction region on the later-arriving second pulse. Therefore, the spectral reshaping is markedly asymmetric with respect to $\Delta t = 0$. As the pressure increases, the coupling between the color components strengthens, and the system transitions from predominantly spectral broadening to a complex $\Delta t$-dependent spectral reshaping with possible suppression of individual spectral regions. The most pronounced reshaping is observed around 400 nm, indicating the high sensitivity of the short-wavelength component to the two-color filamentation dynamics.

A comparison of the gases Ar, $N_2$, and $CO_2$ shows that the nature of the spectral reshaping depends not only on pressure but also on the properties of the nonlinear medium. In Ar (top row in Fig. 2), which has no molecular degrees of freedom, the spectral dynamics can be considered as the simplest case for analyzing the electronic nonlinearity and the plasma contribution. Here, in the spectra near the fundamental frequency, in addition to broadening, a pressure-induced blue shift of the spectral maximum, $\delta\omega < 0$, is clearly observed, which is due to the prevailing negative plasma contribution to the nonlinear phase of the wave $\Phi_{nl}$: $\delta\omega = \partial_t \Phi_{nl} \propto \partial_t \left( n_2 I - \rho_{pl} / 2\rho_{cr} \right)$, where $n_2 I$ is the Kerr contribution proportional to the cubic nonlinear coefficient $n_2$ and the optical intensity $I$, and $\rho_{pl}$ and $\rho_{cr}$ are the free-electron density and the critical plasma density, respectively. In the molecular gases $N_2$ and $CO_2$, more complex spectral structures are observed, which depend not only on the magnitude but also on the sign of the delay, evidently due to the additional molecular degrees of freedom that provide a red-shifted Raman contribution to the effective nonlinear polarizability of the medium [15].

The most intricate behavior is observed for $CO_2$. Already at 6 atm, a pronounced dependence of the spectral transformation on the delay appears, and at 11 atm it becomes even more evident. In the region around 400 nm, a broad spectral structure with distinct suppression features is seen. Meanwhile, the red component near 800 nm also changes, but its structure remains considerably more localized in wavelength. This indicates that the two-color interaction is not reducible to independent broadening of the two original pulses. Instead, spectral energy is redistributed between the components, and the result is determined simultaneously by pressure and time delay. However, quantitative separation of the electronic, molecular, and plasma contributions requires additional experiments. Overall, the results in Fig. 2 demonstrate the feasibility of controlling the supercontinuum spectrum by simultaneous variation of gas pressure and time delay between the two-color pulses.

An interesting feature of the spectrograms in Fig. 2 is the suppression of the blue component at certain delays and elevated pressures. This means that during co-filamentation of the 1ω and 2ω pulses, the field energy is redistributed among different spectral components due to nonlinear phase modulation in the plasma wake already formed by the first pulse. Consequently, enhanced nonlinear interaction does not necessarily imply an increase in supercontinuum intensity across the entire spectral range. Under certain conditions, it can lead to destructive energy redistribution and suppression of a specific spectral region. This demonstrates the possibility not only of broadening the supercontinuum but also of dynamically "switching off" certain portions of its spectrum.

Quantitative information on the spectral transformations of the femtosecond supercontinuum during two-color filamentation in gas is provided by Fig. 3, which shows the root-mean-square (effective) spectral width $\Delta\lambda$ as a function of the interpulse delay. As in Fig. 2, in processing each spectrum, the spectral data were preliminarily calibrated for the spectrometer transmission function and then normalized to the maximum value attained at the given gas pressure. The parameter $\Delta\lambda$ was calculated as the centered second moment of the squared modulus of the spectral field $|U_\lambda|^2$ according to the standard expression:

$$\Delta\lambda = 2\left[\int |U_\lambda|^2 \left(\lambda^2 - \lambda_g^2\right) d\lambda \Big/ \int |U_\lambda|^2 \, d\lambda\right]^{1/2} \qquad (1),$$

where $\lambda_g$ is the spectrum centroid.

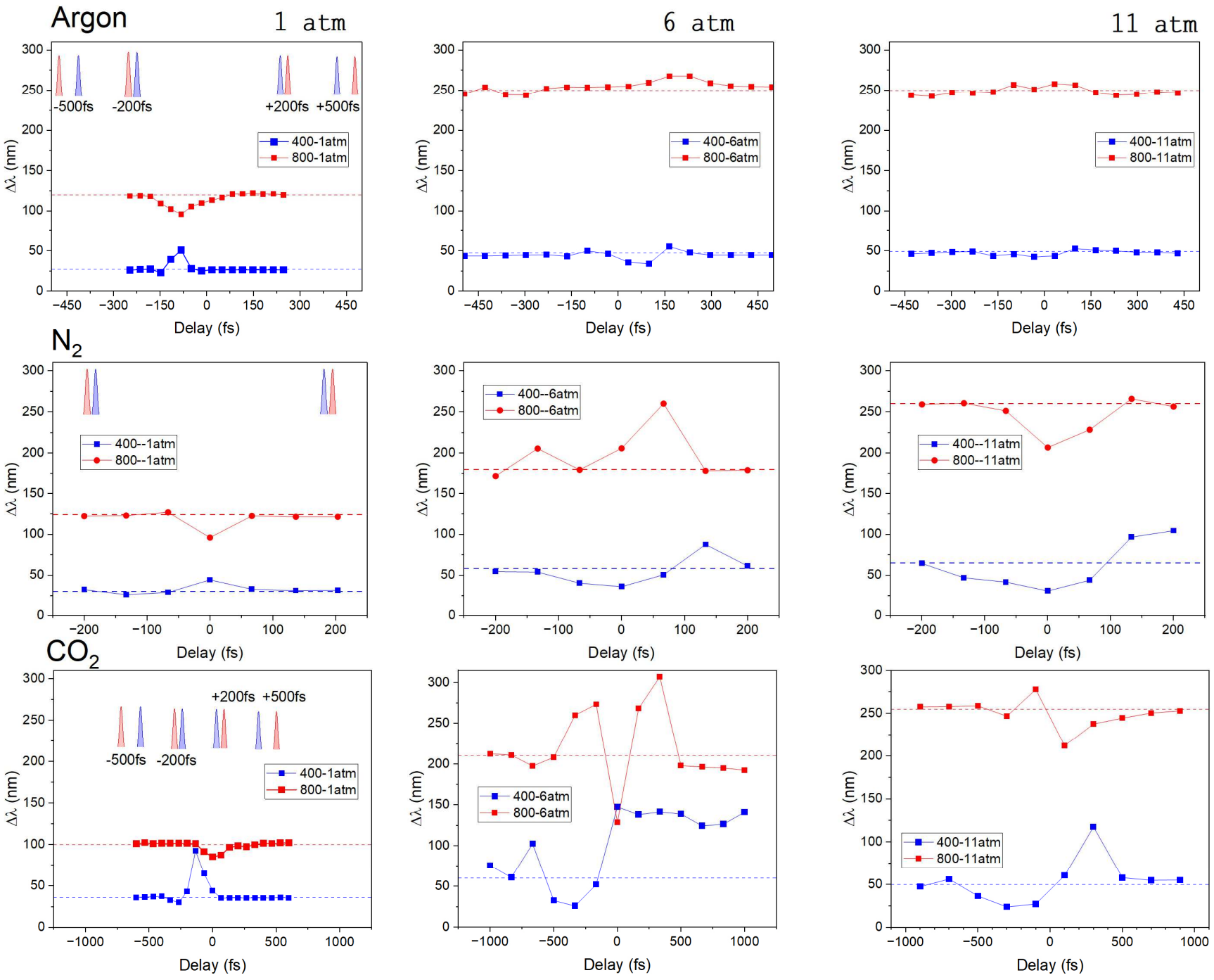

Fig. 3. Root-mean-square (effective) spectral width Δλ as a function of interpulse time delay for different gases and cell pressures (see caption to Fig. 2). Blue and red curves correspond to the spectral components in the 400-nm and 800-nm regions, respectively; the spectral widths of the isolated pulses are indicated by dashed lines.

From the analysis of these dependencies, one can see that, overall, at atmospheric pressure, Δλ varies relatively weak for all investigated gases, indicating a comparatively weak coupling between the two color components. Upon increasing the pressure to 6 atm, the dependence of Δλ on the delay becomes significantly more pronounced, with the strongest changes observed for the short-wavelength component in the 400-nm region. Upon further pressure increase to 11 atm, the pronounced dependence of the supercontinuum spectral shape on Δλ persists and becomes especially noticeable for the molecular gases. Evidently, increasing the pressure shifts the nonlinear system from a regime of weak coupling between the two color components into a regime where the result is determined by their joint dynamics.

Figure 3 also shows that the dependence of the effective spectral width Δλ on the delay has a substantially different character for Ar, $N_2$, and $CO_2$. For argon, which is an atomic gas, an increase in pressure mainly leads to an increase in the average level of spectral broadening, while a pronounced quasi-periodic modulation of Δλ with changing Δ*t* remains relatively weak. At 6 atm, small variations appear for the 800-nm component on a scale of several hundred femtoseconds, but at 11 atm the dependence becomes nearly smooth. For the 400-nm component, the changes are also localized in a relatively narrow delay range. This indicates that in Ar, the dynamics are dominated by fast electronic nonlinear processes and plasma interaction, whereas no pronounced delayed molecular response is present.

A completely different picture is observed in $N_2$. Already at a moderate pressure of 6 atm, a distinct non-monotonic dependence of Δλ on the delay appears, with characteristic changes also occurring on the scale of hundreds of femtoseconds. At transition to 11 atm, the modulation amplitude of the Δλ parameter increases. For example, for the 800-nm component, successive maxima and minima appear, and for the 400-nm component the range of width variation becomes even more substantial. Thus, an increase in pressure not only enhances the spectral reshaping but also leads to a clearer manifestation of its Δt-dependent structure.

This trend is most pronounced in carbon dioxide. In contrast to $N_2$, where the main changes are concentrated roughly within a few hundred femtoseconds, in $CO_2$ the pronounced modulation of the spectral width persists over a much wider delay range, up to about 1 ps and beyond. This is especially evident at 6 atm, when the width of both spectral components exhibits several successive maxima and minima as the delay is varied from negative to positive values. At 11 atm, this oscillatory structure remains, although its shape changes. Consequently, $CO_2$ is characterized not only by a larger amplitude of spectral reshaping but also by a considerably slower dynamics of its formation.

Indeed, during propagation in a molecular gas, an intense laser pulse can excite coherent rotational dynamics of the molecules [15, 16] and give rise to a time-dependent quantum wake generated by the nonadiabatic rotational alignment of gas molecules [4, 5]. As a result, the induced nonlinear polarization of the medium, $\mathbf{P}_{nl}$, is no longer an exclusively instantaneous function of the field intensity, as in an atomic (noble) gas:

$$\mathbf{P}_{nl}(t) = \mathbf{P}_{el}(t) + \mathbf{P}_{rot}(t) + \mathbf{P}_{pl}(t), \tag{2}$$

where $\mathbf{P}_{el}$, $\mathbf{P}_{rot}$, $\mathbf{P}_{pl}$ are instant electronic, retarded rovibrational molecular and free-electrons (plasmonic) response, respectively. For $N_2$ and $CO_2$, it is precisely the rotational contribution $\mathbf{P}_{rot}$ that can potentially account for the appearance of a longer temporal structure in the dependence $\Delta\lambda(\Delta t)$. Moreover, the larger moment of inertia of the $CO_2$ molecule and the greater complexity of its rotational spectrum [17], as compared to the $N_2$ molecule, give rise to the longest sub-picosecond time scales in the dynamics of the rotational response $\mathbf{P}_{rot}$ [13].

Of particular interest is the connection between the observed temporal dynamics of the spectral width and the development of multifilamentation. With increasing gas pressure, the enhancement of nonlinear refraction leads not only to a greater degree of spectral broadening but also to a more complex spatial structure of pulse propagation [13, 14]. Under conditions where the pulse power substantially exceeds the critical self-focusing power, spatial instabilities can give rise to the formation of multiple filaments. In a two-color configuration, this results in an ensemble of interacting filaments generated by the 1ω and 2ω components, whose spatial positions and relative contributions to the resulting spectrum depend on the time delay between the pulses. Consequently, the measured spectrum is the result of the joint contribution of several nonlinear propagation channels. Each filament generates its own nonlinear phase and plasma trail, and in molecular gases a spatially distributed molecular excitation in the form of rotational wave packets also remains [18]. When $\Delta t$ is varied, the second pulse interacts with a different configuration of these nonlinearly modified regions. Therefore, changing the delay can lead not only to a change in the degree of cross-phase modulation but also to a redistribution of energy among the different filamentation channels. As a result, the effective supercontinuum width acquires a non-monotonic, oscillatory dependence on the time delay.

From this perspective, the increase in the amplitude of oscillations in the discussed dependence with increasing pressure is a direct consequence of the enhanced nonlinear dynamics accompanying the formation of multiple filaments and plasma trails. Indeed, as the gas density (pressure) increases, the nonlinear response of the medium grows and the effective critical power for self-focusing decreases, which favors the development of multifilamentation. This leads to an increased sensitivity of the spatial filament configuration to the relative delay of the two pulses. Therefore, in the transition from normal to elevated pressure in Fig. 3, one observes not only an increase in the average spectral width but also a strengthening of its modulation with $\Delta t$.

Taken together, the results presented in Figs. 2 and 3 allow us to interpret the observed spectral dynamics as a manifestation of the spatiotemporal evolution of multifilamentary two-color filamentation. Spectral broadening, its suppression, and the oscillations of the effective width with changing delay are thus not independent effects, but rather different manifestations of the energy redistribution among interacting filaments, their plasma channels, and the cross-modulation wakes originating from the cubic polarizability of the medium. Such complex nonlinear interaction becomes particularly prominent at high pressures, where the spatial inhomogeneity of filamentation is significant, and a simple single-filament model is no longer sufficient to describe the observed spectral dynamics.

## 4. Conclusions

We experimentally investigate the spectral dynamics of femtosecond supercontinuum generation during collinear two-color filamentation of fundamental-frequency (1ω) pulses at 800 nm and second-harmonic (2ω) pulses at 400 nm in Ar, $N_2$, and $CO_2$ gases at pressures up to 11 atm. The main control parameters for the nonlinear interaction are the gas pressure and the relative time delay between the pulses. As shown, an increase in the density (pressure) of the nonlinear medium leads to a qualitative change of the joint supercontinuum character, when the relatively weak interaction between the two spectral components at atmospheric pressure transitions to a regime of pronounced delay-dependent spectral transformation at elevated pressures.

It is established that varying the time delay results not only in a change in the degree of spectral broadening (SC spectral width) but also in a redistribution of energy between the spectral components. At elevated pressures, within certain interpulse delay ranges, a marked reduction in spectral intensity near 400 nm is observed, indicating the possibility of controlled selective suppression of specific portions of the supercontinuum. The observed asymmetry of the spectral reshaping with respect to $\Delta t = 0$ is associated with the different sequence of pulse interaction with the nonlinear medium. The first pulse generates a nonlinear response and a plasma trail, and in molecular gases also a dynamically induced orientational polarization wake. Together, these trails then determine the propagation dynamics of the second pulse. Thus, the spectral response of the medium is governed by the combined action of self-phase modulation, cross-phase modulation of the pulse fields, and the plasma contribution to the nonlinear refraction.

An analysis of the effective spectral width of the two-color SC shows that, at atmospheric pressure, the dependence of the spectral characteristics on the time delay is relatively weak. Upon increasing the pressure to 6 and 11 atm, the amplitude and non-uniformity of the oscillations in the spectral width increase substantially, especially for the short-wavelength component near 400 nm. This indicates a transition from a regime of weak coupling between the color components to a regime of their coupled nonlinear dynamics. Meanwhile, the spectral reshaping is not limited to a monotonic enhancement of broadening. Depending on the delay between the two harmonics, a competition between broadening and selective suppression of spectral regions is observed, reflecting energy redistribution during two-color filamentation.

Notably, substantial differences are found between two-color filamentation in atomic and molecular gases. In Ar, an increase in pressure is mainly accompanied by an increase in the average level of spectral broadening, whereas pronounced modulation with changing $\Delta t$ remains relatively weak. In nitrogen, and especially in carbon dioxide, a considerably more complex temporal structure of the spectral width emerges. For $N_2$, the characteristic changes occur on the scale of hundreds of femtoseconds, whereas in $CO_2$ substantial reshaping persists up to picosecond delays. This is consistent with the presence of a delayed rotational contribution to the cubic nonlinear polarization in molecular gases. The larger moment of inertia and the more complex rotational spectrum of $CO_2$ compared to $N_2$ may be responsible for the longest temporal structure observed in the spectral dynamics. However, quantitative separation of the electronic, molecular, and plasma contributions requires further investigation.

The obtained results should also be considered in the context of the development of multifilamentation with increasing pressure. The enhancement of the nonlinear response of the medium promotes the formation of a spatially inhomogeneous system of interacting filamentation channels, each of which creates its own nonlinear and plasma wake. In the two-color regime, the relative time delay determines the sequence of pulse interaction with these nonlinearly modified regions and thereby influences the energy redistribution among the different filaments of the spectrum. In this sense, the observed oscillations of the effective spectral width can be viewed as a spectral manifestation of the spatiotemporal dynamics of multifilamentary two-color filamentation.

Thus, we demonstrate that the combination of elevated pressure and controlled time delay between the fundamental and second-harmonic pulses provides an efficient means of controlling the spectral characteristics of a two-color femtosecond supercontinuum. By varying the interaction parameters, it is possible not only to enhance spectral broadening but also to induce its suppression, including selective broadening suppression of UV component. The present results expand the understanding of two-color filamentation dynamics in dense gases and highlight the promise of using controlled multifilamentation for the generation and temporal-spectral manipulation of broadband coherent radiation.

**Funding.** Ministry of Science and Higher Education of Russian Federation (IAO SB RAS); project FFUU-2022-0004 of the Institute of Spectroscopy of the RAS

**Conflict of Interest.** The authors have no conflicts to disclose.

**Data availability**. Data underlying the results presented in this paper are not publicly available at this time but may be obtained from the authors upon reasonable request.